\documentclass[prl,twocolumn,preprintnumbers,amsmath,amssymb,superscriptaddress]
{revtex4}

\usepackage[pdftex]{graphicx, color}

\usepackage{amsmath}
\newcommand{\ket}[1]{{| #1 \rangle}}

\usepackage{amsmath}
\usepackage{amssymb}
\usepackage{amsthm}
\usepackage{amsfonts}						%
\usepackage{graphicx}
\usepackage[english]{babel}             	
\usepackage{graphicx}                   	
\usepackage{color}

\begin{document}

\title{Controlling single-photon Fock-state propagation through opaque scattering materials}

\author{Thomas J. Huisman} \thanks{Both authors contributed equally}
\author{Simon R. Huisman} \thanks{Both authors contributed equally}
\author{Allard P. Mosk}
\author{Pepijn W.H. Pinkse}\email{p.w.h.pinkse@utwente.nl, www.adaptivequantumoptics.com}
\address{MESA+ Institute for
Nanotechnology, University of Twente, PO Box 217, 7500 AE Enschede, The Netherlands}

\date{\today}

\begin{abstract}
The control of light scattering is essential in many quantum optical experiments.
Wavefront shaping is a technique used for ultimate control over wave propagation in multiple-scattering materials by adaptive manipulation of incident waves. We control the propagation of single-photon Fock states in opaque scattering materials by phase modulation of the incident wavefront. We enhance the probability that a single photon arrives in a target output mode with a factor 10. Our proof-of-principle experiment shows that wavefront shaping can be applied to non-classical light, with prospective applications in quantum communication and quantum cryptography.
\end{abstract}
\maketitle


Multiple scattering has become an exciting platform for quantum optical experiments \cite{Smolka2009, Bromberg2009, Peruzzo2010, Sapienza2010, Lahini2010, Peeters2010}. In addition, adaptive manipulation techniques controlling light propagation are successfully applied to optimize the desired quantum interference \cite{Bonneau2012, Polycarpou2012}. Non-classical correlations are observed, even for opaque scattering materials \cite{Smolka2009, Peeters2010, Lodahl2005}. Light transport in a multiple-scattering medium can be considered as a linear transformation of a multi-mode system by a scattering matrix, which results generally in a speckle pattern \cite{Sheng1995, Beenakker1997, Akkermans2007}. Each far-field speckle spot is a solid angle that represents a single output mode of the system. These days, parts of the scattering matrix can be measured \cite{Popoff2010}. Adaptive optical techniques make it possible to select elements of the scattering matrix of opaque scattering materials for desired quantum interference.  This opens opportunities for sample characterization, quantum patterning \cite{Boto2000}, secure key generation, or quantum transport through disordered media \cite{Ott2010}.

Wavefront shaping is an adaptive optical technique where one uses spatial light modulation in combination with strongly scattering materials to achieve ultimate control over light in space and time \cite{Vellekoop2007, Aulbach2011, Katz2011, Gjonaj2011, Putten2011, Mosk2012}. Recent wavefront-shaping experiments have transformed opaque media in equivalents of waveguides and lenses \cite{Vellekoop2007, Putten2011} and optical pulse compressors \cite{Aulbach2011, Katz2011} that are inherently robust against disorder and imaging errors. All previously reported wavefront-shaping experiments have been performed with classical light.

\begin{figure}[t!]
  \includegraphics[width=8.3 cm]{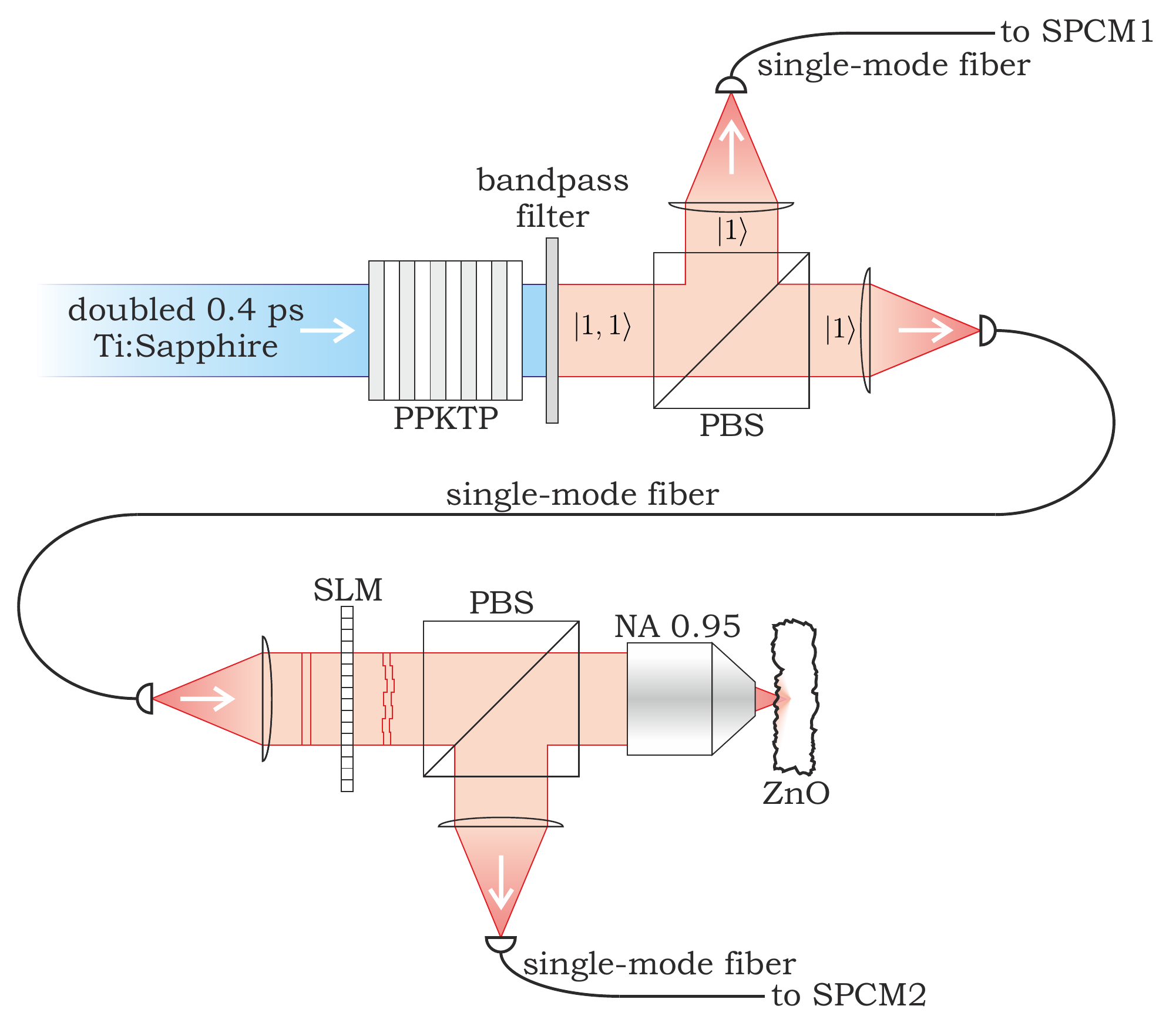}
\caption{\textit{(color online)} Setup for single-photon wavefront shaping. The entangled photon pairs generated in the PPKTP-crystal are separated by a polarizing beam splitter cube (PBS). One photon is fiber-coupled to a single-photon counting module (SPCM1). The conjugate photon is phase-modulated with a spatial light modulator (SLM). The phase-modulated single-photon state is focused on a layer of white paint (ZnO). Only multiple-scattering events are selected by fiber-coupling the reflection of the PBS to SPCM2.}
\label{fig1}
\end{figure}

In this letter we report the first experimental demonstration of wavefront shaping with quantum light. We control light propagation of single-photon Fock states $\ket{1}$ through a layer of white paint. We enhance the probability that a single photon arrives in a target output mode with a factor 10.

A sketch of our setup is shown in Fig. \ref{fig1} and consists of two parts: a quantum-light source (top) and a wavefront-shaping setup (bottom). The quantum-light source is based on Refs. \cite{Huisman2009, Bimbard2010}. A mode-locked Ti:Sapphire laser (Spectra-Physics, Tsunami) emits transform-limited pulses at a repetition rate of $82$ MHz with a pulse width of approximately $0.4$ ps and a center wavelength of $790.0$ nm. Typically $600$ mW was incident on a 5-mm-long BBO nonlinear crystal cut for Type-I frequency doubling. After spectral and spatial filtering, approximately $70$ mW of 395 nm pulses are focused into a 2-mm-long periodically-poled KTP (PPKTP) crystal cut for type-II spontaneous parametric down-conversion (SPDC) in a single-pass configuration. The polarization-entangled output state can be approximated by:
\begin{equation}
\ket{\Psi} \approx \sqrt{(1-\gamma^2)}\ket{0,0}+ \gamma \ket{1,1},
\end{equation}
where the numbers represent the Fock states in the separate polarization modes, and the constant $\gamma \ll 1$ is proportional to the pump field and the effective nonlinear susceptibility of the PPKTP crystal. The modes are separated with a polarizing beam splitter (PBS) and coupled into single-mode fibers. The reflection is used as trigger and the transmission as signal mode. The trigger mode is fiber coupled to a single-photon counting module (SPCM1, Perkin-Elmer SPCM-AQHR-13). Click rates were detected up to $800 \cdot 10^3$ s$^{-1}$, which is state-of-the-art \cite{Huisman2009}. A Hanburry-Brown-Twiss experiment and a Hong-Ou-Mandel experiment confirmed  that a trigger event corresponds to a heralded preparation of $\ket{1}$ in the signal channel with a fidelity of $98 \pm 1 \%$. The signal mode is coupled in a single-mode fiber and directed to the wavefront-shaping setup.

The wavefront-shaping setup is based on \cite{Vellekoop2007}, with the main difference that we use a reflection configuration for practical reasons. The signal mode is incident on a spatial light modulator (SLM, Hamamatsu LCOS-SLM) and focused on a layer of white paint with an objective (NA=0.95). The layer of white paint consists of ZnO powder with a mean free path of $0.7 \pm 0.2$ $\mu$m. The reflected speckle pattern is collected with the same objective. The detected speckles are reflected from a PBS to select only multiple-scattered light. One of the modes can be focused into a single-mode fiber that is connected to SPCM2 (Perkin-Elmer SPCM-AQHR-14). 
For convenience we use transform-limited laser pulses to simulate heralded signal photons during wavefront shaping. Calculations based on the dispersion relation of PPKTP  \cite{Grice1997} indicate that the expected spectral width of the photons is not more than a factor 2 broader than the laser pulses. After optimization with laser light, the same phase pattern is used on the single-photon states. 

\begin{figure}[]
  \includegraphics[width=8.3 cm]{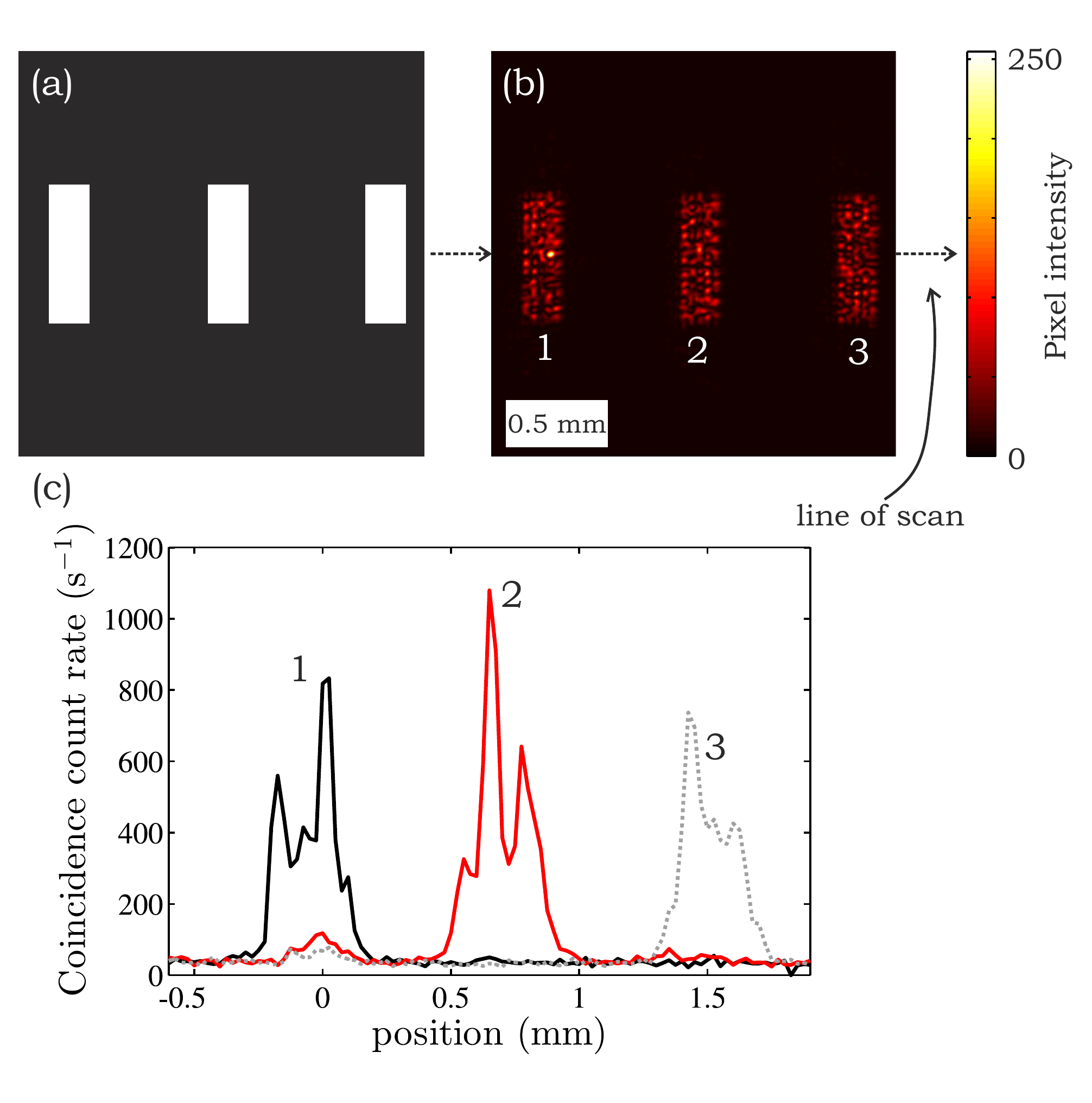}
\caption{\textit{(color online)} Programmable single-photon diffraction patterns. (a) Target diffraction pattern consisting of 3 bars. (b) Measured diffraction pattern imaged on a CCD camera for laser pulses. (c) Measured coincidence count rate for single-photon diffraction when a multimode fiber is scanned over the three bars along the line indicated in (b). Each bar is imaged separately (black, red, grey-dashed).}
\label{fig2}
\end{figure}

Phase modulation of entangled states has been reported with permanent phase masks and SLMs relevant for imaging and communication purposes \cite{Stutz2007, Valencia2007, Cialdi2010, Salakhutdinov2012}. These experiments exploit spatial quantum correlations. Only a single spatial mode of SPDC is modulated in our experiment. In Fig. \ref{fig2} we show an example of programmable diffraction patterns with single photons, where the diffracted beam from the SLM was directly focused on a CCD camera or SPCM2. Fig. \ref{fig2}(a) shows the intended diffraction pattern, consisting of 3 vertical bright lines. Fig. \ref{fig2}(b) shows the diffraction pattern for incident laser pulses imaged on a CCD camera. The 3 lines are clearly visible, where the hologram software (Holoeye) gives a speckle-like intensity distribution of the bars. After we stored the corresponding phase pattern on the SLM, the incident laser pulses were replaced with single-photon states, and the CCD camera was replaced with a multimode fiber with a $62.5$ $\mu$m core diameter connected with SPCM2. The multimode fiber was scanned over the line indicated in Fig. \ref{fig2}(b) to detect the coincidence count rate as a function of position. The results are shown in Fig. \ref{fig2}(c). Each bar was imaged separately, in order to demonstrate the ability to program the diffraction pattern. The coincidence count rate is clearly highest in the intended regions, and also shows speckle-like fluctuations.  The spatial widths of the lines in the coincidence measurements is consistent with the width of the bars convolved with  the multimode fiber.

\begin{figure}[]
  \includegraphics[width=8.3 cm]{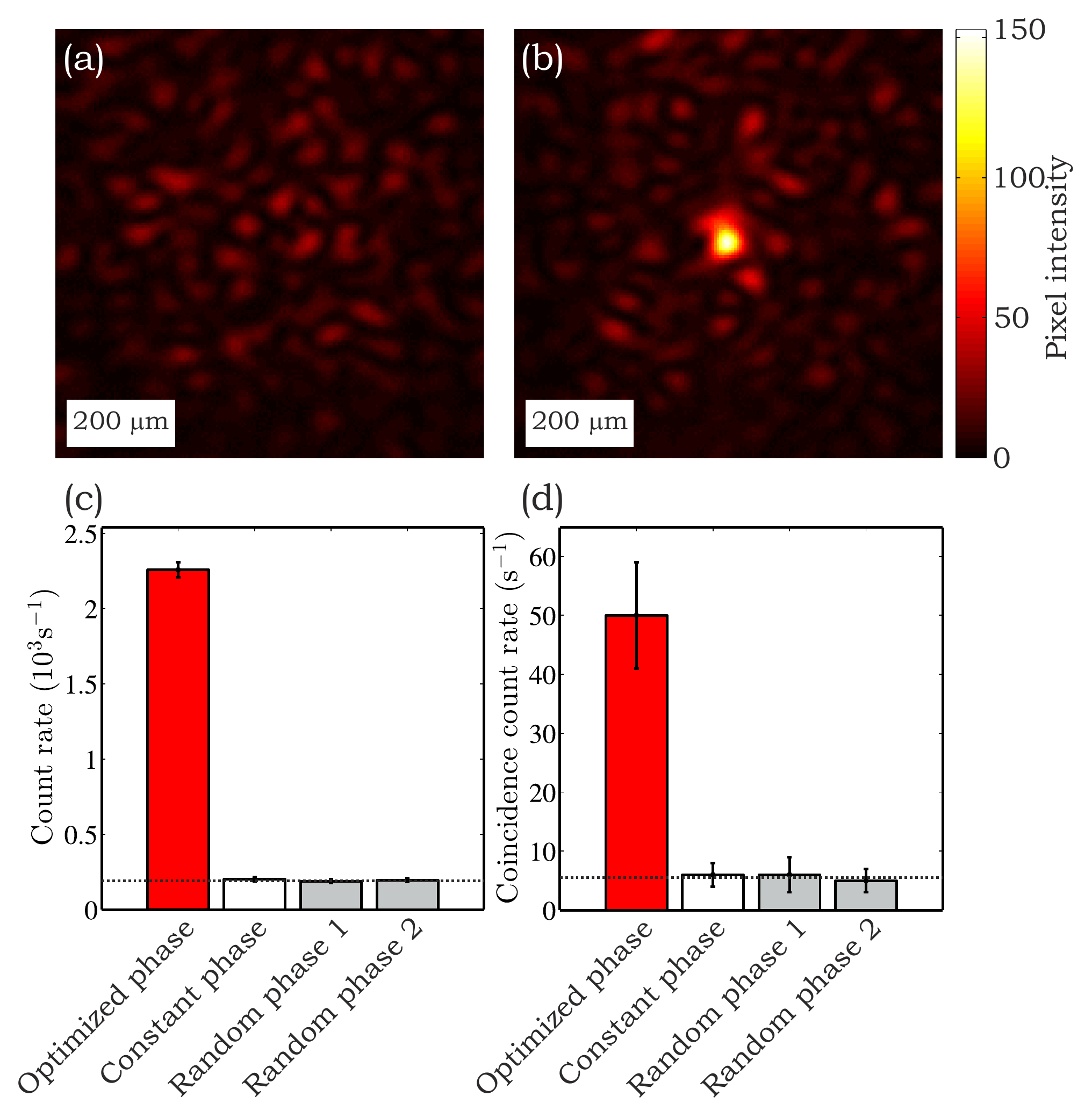}
\caption{\textit{(color online)} Wavefront-shaped single-photon speckle. (a) Speckle pattern imaged on a CCD camera for laser pulses with a random phase pattern. (b) Speckle pattern with a single enhanced mode for an optimized phase pattern. (c) Measured count rate at SPCM2 for an optimized phase pattern (red), constant phase pattern (white) and two random phase patterns (grey). (d) The corresponding coincidence count rate.}
\label{fig3}
\end{figure}

Fig. \ref{fig3} contains the main result of this letter: a wavefront-shaped single-photon speckle pattern. We first use laser pulses to create an optimized classical speckle pattern with a single enhanced mode. SPCM2 was replaced with a Si-pin photodiode to optimize the reflected pattern for classical laser light. The SLM was divided into segments of 20x20 pixels. Approximately 500 segments are used. Each segment is sequentially addressed with a random phase. Only if the intensity increased, this new phase was accepted. This algorithm was repeated approximately 5 times for all segments to obtain a single enhanced mode (approximately $0.5$ hour optimization time). Fig. \ref{fig3}(a) shows the speckle pattern for a random phase pattern and Fig. \ref{fig3}(b) shows the pattern after optimization. The enhancement is limited by the broad spectrum of our laser \cite{Aulbach2011} and laser noise. Nevertheless, the optimized mode has a maximum intensity that is 17.4 times larger than the average intensity. 

Fig. \ref{fig3}(c,d) show the results when the same phase patterns are used with incident single-photon states. Fig. \ref{fig3}(c) presents the count rate on SPCM2 for four different phase patterns; optimized (red), constant (white), and two random (grey). Fig. \ref{fig3}(d) shows the corresponding measured coincidence rate. The horizontal dashed line represents the reference (coincidence) count rate. For the optimized phase pattern we obtained a count rate which is 11.5 times higher than the count rate observed for a constant or random phase pattern. If we take into account the background noise, then this ratio is even 25.3. For the optimized phase pattern, Fig. \ref{fig3}$(d)$, we obtained a coincidence count rate which is 8.9 times higher than the coincidence count rate observed for a constant or random phase pattern. We have, hence, increased the probability for a photon to appear in a desired mode 10-fold, demonstrating the capability of wavefront shaping quantum light.

It is expected that the enhancement can be increased by matching the spectrum of the laser pulses to the SPDC. The enhancement could be improved by orders of magnitude when the temporal width of the single-photon states is increased, because longer random walks in the opaque scattering media can then contribute constructively to the optimized mode. 

In summary, we have controlled single-photon propagation in opaque scattering materials with phase-modulation of the incident light. This opens new opportunities to address elements of the scattering matrix to obtain desired quantum interference. Our results are not only limited to single-photon Fock states, but can be implemented for non-classical light in general.  

We thank C.A.M. Harteveld, J.P. Korterik, and F.G. Segerink for technical support and A.I. Lvovsky, K.-J. Boller, W.L. Vos, A. Lagendijk, and J.L. Herek for discussions and support. This work was supported by FOM.

\pagebreak

\end{document}